# Self-Rolled Multilayer Metasurfaces

*Esteban Bermúdez-Ureña and Ullrich Steiner*

Adolphe Merkle Institute, University of Fribourg, Chemin des Verdiers 4, CH-1700 Fribourg, Switzerland

**Abstract**

Multilayer metasurfaces (MLMs) represent a versatile type of three-dimensional optical metamaterials that could enable ultra-thin and multi-functional photonic components. Herein we demonstrate an approach to readily fabricate MLMs exploiting a thin film self-rolling technique. As opposed to standard layer-by-layer approaches, all the metasurfaces are defined within a single nanopatterning step, significantly reducing fabrication time and costs. We realize two MLMs platforms relying on widely used nanopatterning techniques, namely focused ion-beam and electron-beam lithographies. A first example are MLMs comprised of nanohole patterns structured into metal-dielectric seed bilayers. The second platform is comprised of vertical stacks of angled plasmonic nanorod arrays separated by thin dielectric layers. Such angled MLMs exhibit a selective response to circularly polarized light, in agreement with previous works relying on layer-by-layer processes. Our approach can pave the way for the efficient prototyping of novel MLMs, such as devices with varying number of layers and configurations that can be fabricated on a single chip.

Optical metamaterials currently represent a prominent field in material science and engineering. The possibility to control and manipulate photons in ways not possible with naturally occurring materials (e.g., through negative refractive index)[1] is enabled by engineering metals and dielectrics in subwavelength unit cells within small-footprint media.[2,3] Over the past decade, the development of functional photonic systems integrating metamaterial components has gained considerable momentum.[4,5] Such advances have been facilitated in part by improved and accessible nanofabrication techniques, theoretical frameworks that help to understand and predict new phenomena, as well as available and easy-to-use three-dimensional (3D) optical simulation platforms that aid in the structure design process.

One of the most versatile platforms to explore new functional metamaterial devices are those involving single and multilayer metasurfaces (MLMs),[6–8] by virtue of the structure design freedom associated with the nanopatterning techniques involved. The latter is in contrast, for example, with colloidal[9] or polymer self-assembly[10] template transfer methods, where the geometry of the unit cells in the metamaterials are typically fixed with limited design freedom.

Dielectric metasurfaces have shown great progress towards sensing and imaging applications with the development of so-called flat meta-optics (i.e., structured thin films that allow near-arbitrary phase control), with a strong focus on single layer metasurfaces.[11,12] Furthermore, multilayer configurations with added functionalities (e.g., multi-wavelength operation and polarization control) have also emerged in recent years.[13] On the other hand, single and multilayer plasmonic metasurfaces can provide a diverse playground to exploit novel non-linear phenomena.[5] One example is the field of optical chirality,[14] where the strong circular dichroism exhibited by the structures can be exploited in applications such as high-sensitivity sensing and differentiation of chiral elements,[15,16] as well as in ultra-thin circular polarizers.[17]

In the particular case of MLMs, the common approach to fabricate such devices has relied on layer-by-layer processes,[18–21] wherein each layer requires precise planarization steps apart from the standard nanopatterning and material deposition/etching steps involved in generating each metasurface. For certain applications, and especially in exploratory research phases, such approach is not only time and cost consuming, but can also be subject to issues with the layer-to-layer uniformity and reproducibility, given the multiple steps involved.

Here, we demonstrate a fabrication approach that greatly reduces the number of steps required to build MLMs devices. Our approach exploits a self-assembly method known as the thin film self-rolling technique,[22,23] where layers exhibiting strain gradients stemming from their manufacture across the film thickness, self-roll upwards or downwards depending on the sign of the strain gradient[24] when selectively releasing them from a substrate by removing a sacrificial under-layer. The result is a microtubular structure with multilayered surface walls according to the total number of turns completed.[25,26] Up to now, this technique has been exploited to produce a wide range of self-rolled devices covering a variety of fields, including electronic[27,28] and magneto-electronic devices,[29] self-propelled micro-engines,[30] biological in-vitro platforms,[31,32] as well as a various integrated photonic systems based on optical microtube resonators.[33–35] Self-rolled metamaterials have also been explored,[36] mainly focused on hyperbolic multilayers, for example for sub-wavelength imaging[37,38] and emission enhancement in quantum emitters.[39] The concept of stacking nanostructured optical metasurfaces exploiting the self-rolling technique to create MLMs has only been roughly explored experimentally in a system involving uniformly porous metallic films,[40] and theoretically to enable three-dimensional photonic crystals.[41] However, the deterministic fabrication of distinct optical metasurfaces into a multilayer assembly has not yet been experimentally demonstrated.

Figure 1a illustrates the basic concept of our MLMs fabrication approach. Instead of relying on multiple fabrication cycles to achieve the multilayer device, a two-dimensional (2D) metasurfaces area is designed, such that lateral feature elements superpose during the rolling process. Note that this requires only one single nanopatterning step. Being able to control the stress within the film and thereby the roll diameter, an in-plane arrangement of metasurface features is converted into a vertical stacked morphology upon the selective removal of the sacrificial layer. This *single-shot* approach significantly reduces the fabrication steps of the MLMs, particularly for rolls featuring a large number of layers. In addition, since the rolling areas can be pre-defined (e.g., during a photolithography step), and many rolls can be created on a single sample, the number of layers in the rolled-up stack (i.e., number of microtube turns) is readily controlled, and enables the fabrication of devices with varying number of metasurfaces on the same sample. The latter can be quite powerful in the exploratory phase of a particular MLM system, where a large parameter space can be covered by lithographically patterning on single sample.

To demonstrate this method, we have implemented two MLM approaches, employing two standard nanolithography techniques, namely focused ion-beam (FIB) milling and electron-beam lithography (EBL). The two initial processing steps are common to both approaches. They comprise the deposition of a Ge sacrificial pattern (Figure 1b) and a $SiO_2$ pattern aligned atop the sacrificial layer (Figure 1c). Both areas are patterned by photolithography, thin film deposition and lift-off steps. The details of these and further fabrication steps can be found in the Methods section. Figure 1d shows the main steps for the FIB based approach. First, a Au film is deposited on top of the $SiO_2$ layer to create a metal-dielectric bilayer and followed by the FIB milling of the metasurfaces (e.g., nanostructured hole arrays). The samples are then immersed in a $H_2O_2$ solution that selectively etches the Ge layer[25,27] and enables the nanostructured bilayer to roll-up, such that the patterned nanohole arrays are stacked at the top surface of the microtube. In the second approach (Figure 1e), after the patterning of the Ge and SiO$_2$ layer, alignment marks are first etched onto the Si substrate, which are required for the EBL step. The samples are then coated with an electron-beam sensitive resist, which is EBL-patterned, followed by Au deposition and lift-off. In this double-layer system, upon the selective etching of the Ge layer, the film rolls-down to form the microtube structure with stacked nano-feature arrays embedded in the microtube walls.

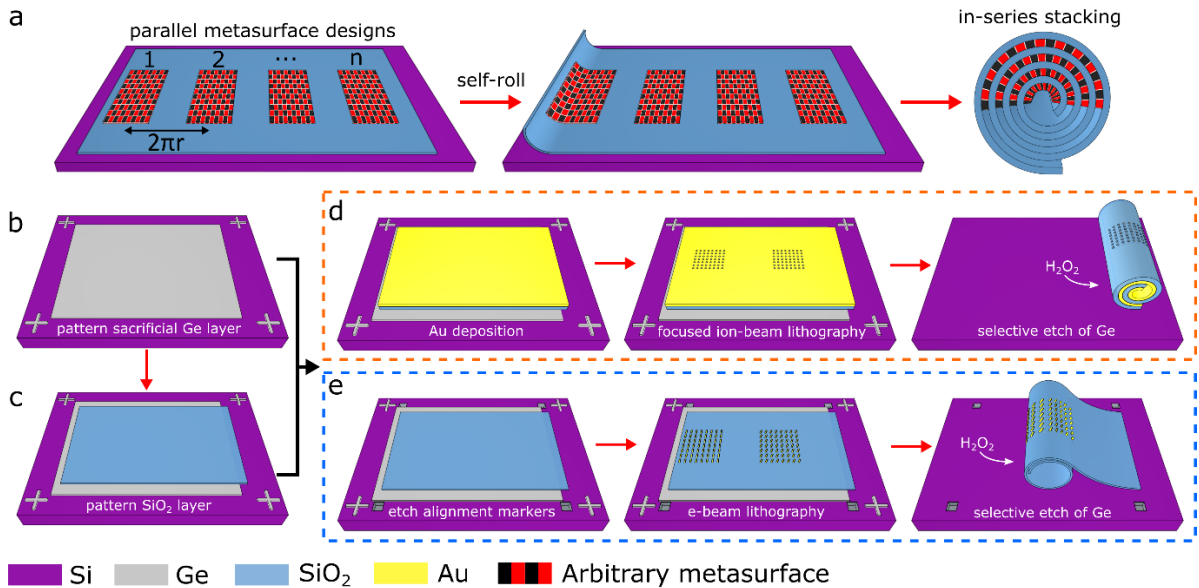

**Figure 1.** Multilayer metasurface (MLM) fabrication. (a) Sketch illustrating the evolution from the design of independent 2D metasurfaces to a multilayer stack exploiting a *single-shot* thin film self-rolling mechanism. (b-c) Initial steps for the two approaches in (d-e). First, a Ge sacrificial pattern and alignment marks are defined by photolithography, deposition and lift-off. A second photolithography step defines a rolling area aligned with respect to the sacrificial layer, followed by deposition and lift-off of a SiO$_2$ film. (d) Focused ion-beam (FIB) lithography-based approach: Au is deposited to form a metal-dielectric bilayer, followed by FIB structuring at selected locations. Selectively etching the sacrificial layer enables the self-rolling and stacking of the nanostructured bilayer pattern. (e) Electron-beam lithography (EBL) based approach: Deposition of EBL alignment marks by photolithography followed by dry-etching of the Si substrate, the definition of an array of nano-features by EBL, Au deposition and lift-off. The selective removal of the sacrificial layer causes roll formation and the stacking of the nano-features. In both approaches, the final device is a self-rolled microtube with nanostructured metasurfaces stacked within the top microtube walls.

Results of the FIB-based approach are shown in Figure 2, where top-view scanning electron microscopy (SEM) images show two Au (20 nm)/SiO$_2$(60 nm) bilayers on top of Ge (20 nm), with FIB-milled nanostructured hole arrays prior to the self-rolling step (Figure 2a-b). In Figure 2a, two equal nanohole arrays, with 300 nm diameters and a 700 nm pitch (see close-up inset) were patterned along the rolling direction. Similarly, Figure 2b shows two nanoslit arrays of 240 nm long and 110 nm wide slits and a pitch of 500 nm. The slit orientation of the two patterned regions was orthogonal (see insets). After the selective removal of the Ge layer, the structured bilayers self-rolled into microtubules (side-view SEM images in Figures 2c-d). The roll-up behavior is indicative of a positive strain gradient in the seed bilayer.[24] The multilayer arrangement of the nanohole arrays is revealed by imaging the top surface of the microtubes (Figures 2e-f). In both cases, the top nanohole pattern is distinguished by the darker contrast with respect to the bottom layer, where electrons must penetrate through the top bilayer.

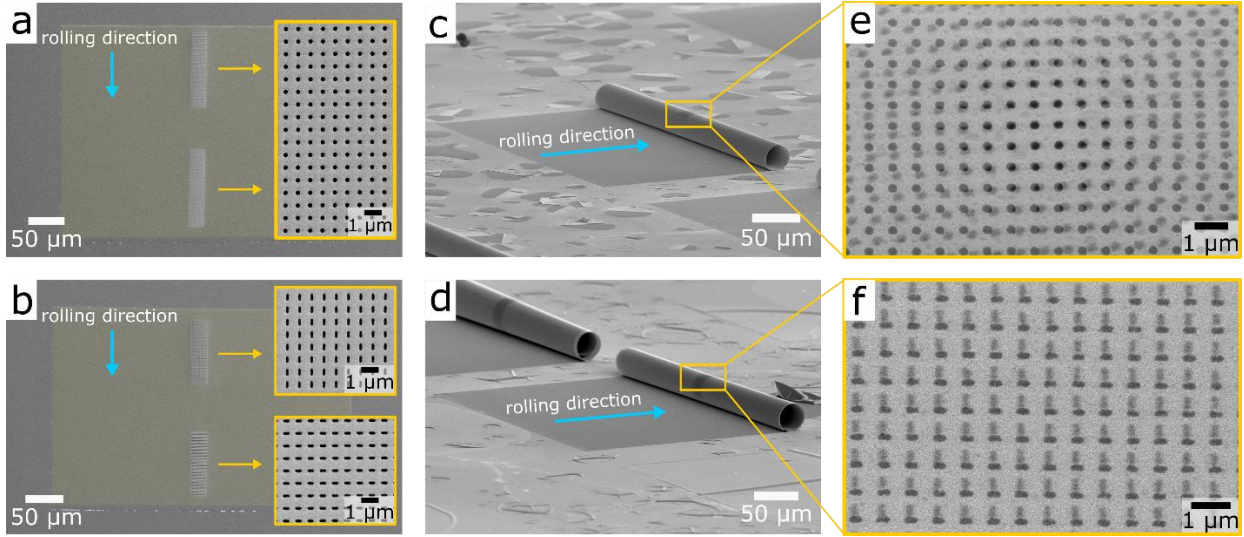

**Figure 2.** Demonstration of FIB-based MLMs. (a-b) Top-view scanning electron microscopy (SEM) images of Au/SiO$_2$ bilayers on Ge sacrificial patterns (the false-color depicts the patterned area). In both cases, two nanohole arrays were milled along the rolling direction, (a) two equal nanohole arrays and (b) two orthogonal nanoslit arrays, shown in the insets. (c-d) Side-view SEM images of self-rolled microtubes after the sacrificial layer removal. (e-f) Top-view SEM images of the rolls in (c-d) respectively, revealing the two nanohole array stacks in the microtube walls.

To demonstrate the second approach, employing EBL nanostructuring, we manufactured a system which was previously demonstrated using a layer-by-layer method,[19] namely the helicoidal stacking of plasmonic nanorods (NRs), which have potential as chiral plasmonic sensors or filters[14] and are therefore referred-to as twisted optical metamaterials.[15,20,21] Figure 3a shows an example of a SiO$_2$(60 nm)/Ge(20 nm) pattern that was deposited onto a Si substrate, and within it, 35 nm thick Au NR arrays (110 nm long, 60 nm wide and 350 nm pitch) which were positioned along the rolling direction. The NRs had 0° and 30° orientations with respect to the microtube long axis (see insets). The six patterned areas with varying distances between the array-pairs allow for a variation of the microtube diameter. After the selective removal of the sacrificial layer, the SiO$_2$ pattern self-rolled into a microtube (Figure 3b) with a diameter of roughly 26 μm, which corresponds closely to the array separation of the two patterned areas in the middle of Figure 3a. The roll-down behavior in this second approach is indicative of a negative strain gradient in the $SiO_2$ layer.[24] The two-layer stack of NR arrays at the top microtube surface is shown in a close-up SEM image (inset in Figure 3b). Within the same sample, self-rolled structures with angled NR array stacks consisting of up to four layers were fabricated, where the NRs were oriented with 30° relative angles between consecutive layers (see the top-view SEM images in Figures 3c-f).

The close-up images in Figures 2e-f and Figures 3c-f also show a current limitation of the self-rolling approach, in particular the limited in-plane alignment of the nanostructures from one layer to the next. In contrast to multiple EBL steps, where alignment between layers can be controlled reproducibly down to a few nanometers, small deviations in the self-rolling process such as tube diameter variations or misaligned rolling directions, can lead to misalignments between each metasurface, in our case up to a couple of microns. Although this can be critical for devices relying on 3D structures that are sensitive to the precise layer-to-layer alignment, as in the case of some plasmonic oligomers,[42] other systems, such as twisted NR arrays are more tolerant to some misalignment between layers.[43] In such periodic metasurfaces, pattern-to-pattern misalignments can be reduced by designing arrays with smaller separations between the nano-features. For applications relying on precise vertical alignment between individual elements, we expect that by carefully engineering the self-rolling protocols and implementing methods such as magnetic-field assisted rolling[44] or the recently proposed capillary induced self-aligning process,[45] will improve the precision of stacking-by-rolling.

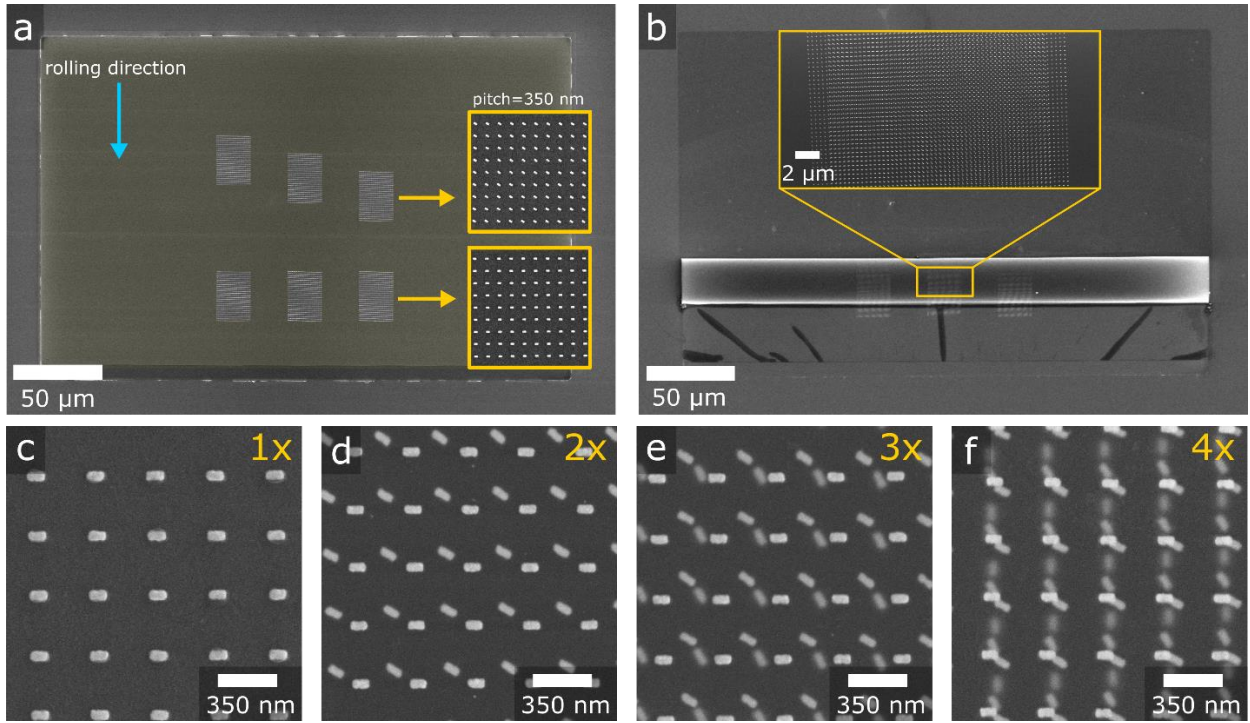

**Figure 3.** EBL-based MLMs. (a) Top-view SEM image of a SiO$_2$ film on a Ge sacrificial layer, indicated by the false-color. NR arrays with 0° and 30° orientations of the rods with respect to the microtube long axis (insets). The six patterned areas with varying distances between the array-pairs allow for diameter variations of the rolls. (b) Top-view SEM image of a self-rolled microtube from the pattern design in (a). The inset shows a close-up SEM image of the central array-pair in (a) showing their overlap. (c-d) Top-view SEM images of microtubes with 1-4 layers of twisting NR array stacks, respectively. Each NR array has a 30° relative angle with respect to the previous layer.

Following the MLM manufacture, their chiral plasmonic properties were investigated by optical reflection measurements. The structured top surface of the rolls was illuminated by circularly polarized light, containing two-layer Au NR stacks with left- and right-handed twist axes. First, the reflection spectra under white linearly polarized (LP) illumination was measured for the case of a single Au NR array onto a SiO$_2$ film (Figure 4a). The black and red curves correspond to incident LP parallel and perpendicular to the NR long axis, respectively. The dashed lines centered around 590 nm and 740 nm, correspond to the transversal and longitudinal localized plasmon resonances, as expected from finite-domain time-difference simulations (see the Supporting Information).

The circular polarization (CP) response was determined after the self-rolling step. Figure 4b shows the CP reflection measurement of a two-layer stack of Au NRs with a left-handed twist (inset). The blue and yellow curves correspond respectively to the reflection of left- and right-handed CP (LCP & RCP) light incident onto the patterned microtube surfaces. Figure 4c shows the differential CP reflection ($R_{LCP}$-$R_{RCP}$), analogous to the way circular dichroism is typically presented in transmission measurements.[20] The sign of the differential reflection is inverted for the two different relative handedness, as expected, and both plots exhibit crossings near the measured plasmon resonances (Figure 4a). The difference in the differential CP reflections for the two twisted NR stacks are attributed to the variation of NR alignment from one layer to the next (insets in Fig.4c). Nevertheless, despite the off-axis misalignment in the right-handed structure, the signs of the curves are not affected by this.[43] The latter is understandable given that a stack with 30° relative angles does not change its handedness as a function of the relative in-plane positions, as would be the case for NRs with 90° relative angles or other 3D oligomer designs.[42] We also note that the uniform color observed through optical microscopy imaging (see the Supporting Information) suggests we can expect a low variation of the optical response as a function of the measurement position within the overlapping region, due to the periodicity of our structures. In other MLMs configurations that would rely on more precise pattern-to-pattern alignments, more in-depth optical characterizations such as optical mappings with scanning elements could be implemented.[46]

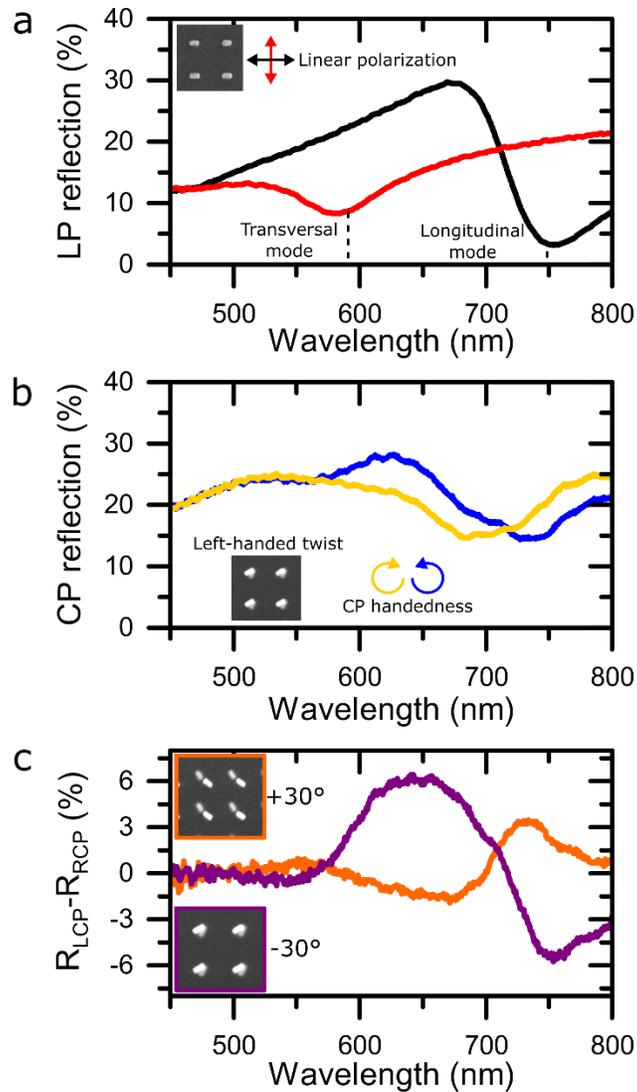

**Figure 4.** Optical characterization of NR based MLMs. (a) Optical reflection spectra of a single layer of NRs on a SiO$_2$/Ge covered Si substrate, for parallel and perpendicularly polarized incident light with respect to the NR long axis (black and red traces, respectively). The position of the transversal (590 nm) and longitudinal (740 nm) modes of the NRs are identified. (b) Optical reflection spectra of a two-layer stack of left-handed stacked NRs under left-handed (blue) and right-handed (yellow) CP illumination. (c) Differential CP reflection of two-layer stacks for left-handed (purple) and right-handed (orange) angled NRs. The handedness of NR-stacking determines the sign of the response.

**CONCLUSIONS**

In conclusion, we experimentally demonstrate a platform for the fabrication of MLMs that relies on a thin film self-rolling mechanism, significantly reducing the fabrications steps involved in standard layer-by-layer approaches.[19–21,42] We demonstrate the versatility of this approach by fabricating structures implementing two of the most common metasurface nanopatterning techniques, namely EBL and FIB milling. This approach should additionally be compatible with other nanopatterning techniques and self-rolling protocols. We show stacks of nanohole arrays structured by FIB milling of metal-dielectric bilayers. Stacking by self-rolling ensures that the nanohole metasurfaces retain their air filling across the multilayer, which is hard to achieve with layer-by-layer methods that rely on planarization steps. Such structured metal-dielectric multilayers could enable novel device concepts in the field of hyperbolic metamaterials.[47] In a second approach, we show MLM-stacking of EBL-produced plasmonic nanostructures, demonstrating up to four-layer stacks of twisted NR arrays. Optical characterization of bilayers with left- and right-handed angled NR arrays reveal their selective response to CP light, in agreement with similar previously reported systems made by layer-by-layer processes.[20,43] A more detailed study of the chiral reflectivity response in devices with larger number of layers and varying nanoparticle designs will be the scope of a future work. Although the stacks presented here are limited to small surface areas, our concept of parallel to in-series MLMs assembly could be implemented on a larger scale by relying on mechanically assisted thin film rolling or folding technologies, which have already been used to explore other multilayered metamaterial configurations.[48,49]

In our approach, the precise control over the number of layers is enabled by photo-lithographically designed rolling areas, rendering this method powerful for prototyping as it facilitates the batch fabrication of devices with varying number of layers on a single substrate. This self-rolling approach should be compatible with the integration of dielectric resonators, quantum emitters or 2D materials, enabling novel hybrid multilayered metamaterial devices that are otherwise difficult to realize with standard layer-by-layer approaches. Furthermore, we foresee the possibility of integrating MLMs in electrical circuitry[27,28] to achieve tunable devices, as well as the implementation of the MLMs in microfluidic platforms,[35] for sensing applications.

**METHODS**

**Structure fabrication.** Photoresist bi-layers (MicroChem LOR 5A –AZ1512 HS) were spin-coated (Süss ACS200 Gen3) onto silicon wafers, exposed with a laser writer (Heidelberg MLA150) and developed (Süss ACS200 Gen3) with AZ 726 MIF (Microchem). The wafers were then coated with a 20 nm Ge layer by means of electron-beam evaporation (Leybold Optics LAB600H), and lift-off was carried out by immersion in Remover 1165 (Microposit), followed by rinsing in IPA and $H_2O$, and a final blow-dry with $N_2$. The coating, exposure and development steps were then repeated to define the second pattern aligned to the Ge sacrificial patterns. A 60 nm $SiO_2$ film was sputter-deposited (Pfeiffer SPIDER600) and followed by a lift-off procedure.

In the case of the process flow shown in Figure 1d of the main text, a 20 nm Au layer, including a 1 nm Cr adhesion layer was deposited at a 60° angle with respect to the source (Moorfield MiniLab80). The angled deposition has a double function; on the one hand, the shadow created at one side of the pattern provided an opening window for the etching of the sacrificial layer, while the coverage on the opposite end of the patterns serves to anchor the structures after rolling.[26] The nanohole arrays were patterned by means of a focused-ion beam (Thermo Scientific Scios 2 Dual Beam). The self-rolling was carried out by selectively etching the Ge sacrificial layer in a 35%

H₂O₂ solution for 3h. The samples were then sequentially rinsed in H₂O and IPA, and finally blow-dried with a N$_2$ gun.

In the case of the process flow shown in Figure 1e of the main text, alignment markers were etched into the Si wafers, which were necessary for electro-beam lithography (EBL). To this end, a 2 μm thick AZ1512 resist was spin-coated onto the sample, followed by exposure and development. 2 μm deep holes were then then dry etched (Alcatel AMS 200 SE) into the substrate, followed by oxygen plasma cleaning (Tepla GiGAbatch) of the remaining resist. A double-layer of electron-beam resists (Microchem MMA EL6 and PMMA 495 2% ), was spin-coated at 6000 rpm onto the substrate and baked at 180 °C for 5 min. Exposure of the nanostructured patterns within the rolling areas was performed with an automatic alignment protocol (Raith EBPG5000+ at 100keV and 650 μC/cm$^2$ dose). The sample was then developed in MIBK:IPA (1:3) for 60s followed by another 60s in IPA and finally dried with a N$_2$ gun. The self-rolling step was the same as described above.

**Scanning Electron Microscopy**. Scanning electron images were taken on a TESCAN Mira3 LM FE and a Thermo Scientific Scios 2 Dual Beam, both operated at 10kV.

**Atomic force microscopy**. The thin film thicknesses were measured with a JPK Nanowizard 2 microscope operated in tapping mode using standard silicon probes (NanoAndMore).

**Optical reflection spectroscopy**. Reflection spectra under linear and circular polarization were carried out in a Zeiss Axio Scope.A1 bright field microscope. Spectra were recorded with an Ocean Optics spectrometer (QE Pro) after collecting light through a 20x objective (Zeiss EC Epiplan-NEO-FLUAR) and 50 μm core optical fiber (Ocean Optics QP-50-2-UV-BX). For such a combination, our collection spot was ~4 μm in diameter. The linear polarization measurements were performed with a Zeiss slider polarizer (427710-9000-000), while the circular polarization states were generated by means of a broadband linear polarizer (Thorlabs WP25M-UB) followed by a super-achromatic quarter waveplate (Thorlabs SAQWP05M-700).

**Optical simulations**. Simulations of the nanorod array response to linearly polarized light were performed by means of a finite-difference time-domain (FDTD) method (FDTD Solutions from Lumerical Inc.).

## ASSOCIATED CONTENT

**Supporting Information**

FDTD optical simulations of the plasmonic nanorod array response to linearly polarized light, and optical image of the microtube with a two-layer nanoparticle pattern stack.

## AUTHOR INFORMATION

**Corresponding Authors**

*E-mail: esteban.bermudez@unifr.ch

*E-mail: ullrich.steiner@unifr.ch

**ORCID**

Esteban Bermúdez-Ureña: 0000-0002-8964-9660

Ullrich Steiner: 0000-0001-5936-339X

**Author Contributions**

E.B-U conceived and developed the fabrication approach, characterized the samples and analyzed the data. Both authors contributed to the manuscript preparation.

**ACKNOWLEDGMENTS**

We thank Bodo Wilts for feedback on the manuscript, and the Center for MicroNanotechnology (CMi) of the École Polytechnique Federal Lausanne (EPFL) for technical support and access to their cleanroom facility. This project has received funding from the European Union's Horizon 2020 research and innovation programme under the Marie Skłodowska-Curie grant agreement No 741855 (E.B-U), from a Swiss Government Excellence Fellowship (E.B-U.), from the Swiss National Science Foundation (Grant No. 163220) and from the Adolphe Merkle Foundation.

**Supporting Information**

# Self-Rolled Multilayer Metasurfaces

*Esteban Bermúdez-Ureña\* and Ullrich Steiner\**

Adolphe Merkle Institute, University of Fribourg, Chemin des Verdiers 4, CH-1700 Fribourg, Switzerland

\*E-mail: esteban.bermudez@unifr.ch, ullrich.steiner@unifr.ch

Pages: S1-S3, Figures: S1, S2.

1. Simulations of the nanorod array illuminated by linearly polarized light

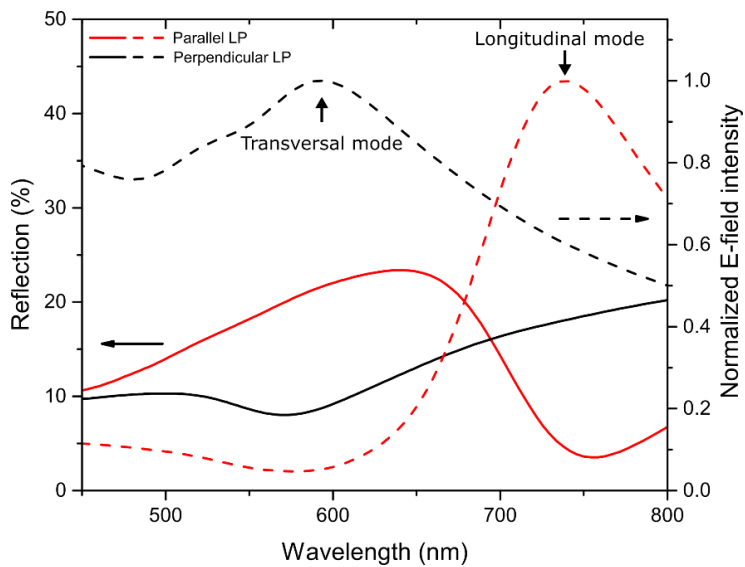

**Figure S5**. FDTD simulations of the optical response of a Au NR array on a $SiO_2$ (60 nm)/Si substrate illuminated by linearly polarized light. Reflection spectra (solid lines) and normalized electric field intensities near the NR edges (dashed lines), for perpendicular incident light with linear polarization aligned parallel (red) and perpendicular (black) to the NR long axis. Based on the electric field response, the transversal and longitudinal modes are identified.

2. Optical microscopy of the angled nanoparticle stacks

In measuring the differential reflection of our stacked layers, we effectively probed average responses, as we are dealing with periodic nanoparticle arrays (pitch of 350 nm) and collecting broadband light from a ~4 µm diameter spot. This corresponds to about 95 nanoparticles at each layer contributing to the measured signal. This is confirmed by optical microscopy imaging (SI Figure 2). Here, even though a lateral misalignment of a few microns was obtained after rolling, the overlapping region exhibited a uniform dark blue color, suggesting that varying the position of our 4 μm collection spot within this area would've yielded similar results to those presented in Figure 4 of the main text.

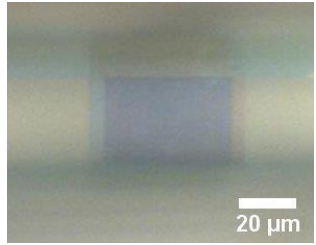

**Figure S6**. Optical microscopy image of a self-rolled microtube with two stacked nanorod arrays. The dark blue area corresponds to the overlapping region between the two layers.